# Amazon Books Rating prediction & Recommendation Model

Hsiu-Ping Lin, Suman Chauhan, Yougender Chauhan, Nagender Chauhan, Jongwook Woo
Department of Information Systems, California State University, Los Angeles
{hlin54, schauha7, ychauha4, nchauha5, jwoo5}@calstatela.edu

**Abstract:** This paper uses the dataset of Amazon to predict the books ratings listed on Amazon website. As part of this project, we predicted the ratings of the books, and also built a recommendation cluster. This recommendation cluster provides the recommended books based on the column's values from dataset, for instance, category, description, author, price, reviews etc. This paper provides a flow of handling big data files, data engineering, building models and providing predictions. The models predict book ratings column using various PySpark Machine Learning APIs. Additionally, we used hyper-parameters and parameters tuning. Also, Cross Validation and TrainValidationSplit were used for generalization. Finally, we performed a comparison between Binary Classification and Multiclass Classification in their accuracies. We converted our label from multiclass to binary to see if we could find any difference between the two classifications. As a result, we found out that we get higher accuracy in binary classification than in multiclass classification.

## 1. Introduction

This paper uses PySpark Machine Learning Models to predict the ratings for books using the Amazon book review dataset. Amazon was originally founded by Jeff Bezos in 1994 and has grown rapidly to become one of the most successful e-commerce businesses in the world. At a rapid rate, Amazon has expanded in the world and has become one of the most popular retailing websites in the world. The success is mainly due to its customer friendly website interface and innovative tools that aid the customers such as providing lists of best sellers, popular books, and the recommendation system. Reviews are generated in the corresponding product when the customers leave their feedback and rating on the website.

The dataset consists of two files Books_rating.csv & books_data.csv Books_rating.csv has information about 3M book reviews for 212,404 unique book and users who gives these reviews for each book. This file also consists of columns such as Id, Title, price, profilename, review/summary, review/text, review/helpfulness, review/score etc. Books_data.csv has details of 212,404 books such as genres, authors, cover, description etc. with columns consisting of such as Title, description, authors, publisher, categories etc.

When a user posts a review on amazon.com, they have the option to post the review text and summary text. Review text, as the name suggests, is an elaborate review typically ranging from 1-2 paragraphs whereas review summary text is a crisp description ranging between 1-3 sentences.

## 2. Related Work

Nicholas et al. [1] discussed the application of various supervised and unsupervised machine learning models to predict rating values for Amazon product reviews. The goal is to create a versatile and accurate model that can handle a wide range of mixed and polarized sentiments expressed in reviews. They explored different embeddings, such as BERT, Word2Vec, Bag of Words, and TF-IDF, and evaluates their impact on model performance. The models include supervised boosting models from Light GBM and CatBoost, as well as deep learning networks. The article covers the dataset curation, preprocessing, and feature engineering process, followed by a detailed analysis of each model's performance. The limitations of the project are also acknowledged, and potential applications for sentiment analysis are discussed, including auto-generated suggestions for rating sentiment and falsified review detection. They concluded with overall results, key findings, and recommendations for future improvement and research.

Petra et al. [2] illustrated investigation combines deep learning and conformal prediction to achieve accurate sentiment analysis of Amazon product reviews across 12 categories. They demonstrated the effectiveness of using machine learning, particularly deep learning, in sentiment analysis, which has become increasingly important in analyzing user opinions. Conformal prediction addresses the lack of confidence measures in machine learning predictions, providing instance-level confidence estimates. The research highlights the generalizability of the approach across different product review categories and its ability to handle imbalanced sentiment classes. The study analyzes highly imbalanced sentiment classes in Amazon customer reviews using deep learning and Mondrian conformal prediction.

Martin et al. [3] discussed the challenge of predicting numerical ratings from text reviews and its significance in consumer decision-making. The authors explore supervised machine learning techniques, including binary and multi-class classification, and logistic regression. They evaluate the performance of state-of-the-art classifiers using datasets from Amazon, addressing the issue of class imbalance through sampling techniques. The results highlight the effectiveness of Naïve Bayes and SVM classifiers, with implications for automating numerical feedback in various contexts. The study emphasizes the importance of star ratings as a quick reference for product quality and provides insights into improving consumer decision processes. Overall, the research contributes to the field of sentiment analysis and suggests avenues for future improvements.

### 3. Specifications

The dataset comprises of the dataset consists of two files Books_rating.csv & books_data.csv. The size of the dataset is 2.9GB. Amazon review Dataset contains product reviews and metadata from Amazon, including 142.8 million reviews spanning May 1996 - July 2014.

Table 1 shows files and size of the files from dataset.

*Table 1 Data Specification*

| Data Set | Size (Total 2.9GB) |
|---|---|
| Books_rating.csv | 2.8 GB |
| books_data.csv | 181.35 MB |

The below table 2 shows the specifications for Hadoop cluster and Spark specifications.

*Table 2 Specification*

| Number of nodes | 3 |
|---|---|
| Hadoop Cluster Version | 3.1.2 |
| Spark | 3.0.2 |
| CPU speed | 1995.309 MHz, 4 core |
| CPU | |
| Memory | 390.7 GB |

### 4. Architecture

The Project architecture is illustrated below in (Figure 1). The Data Source being the first phase, we downloaded the dataset from Kaggle [4]. Then the second phase was Data and Feature Engineering. We used Databricks, zeppelin, Spark CLI, VectorAssemblers, Indexers, Pipeline and Hadoop HDFS. The third phase is Data Modelling which we achieved by using PySpark ML lib and tuning parameters with CrossValidation & TrainValidationSplit etc. Finally, for the Data Validation we used binary and multiclass evaluation, estimators feature importance. The overall Architecture is illustrated below in Figure 1.

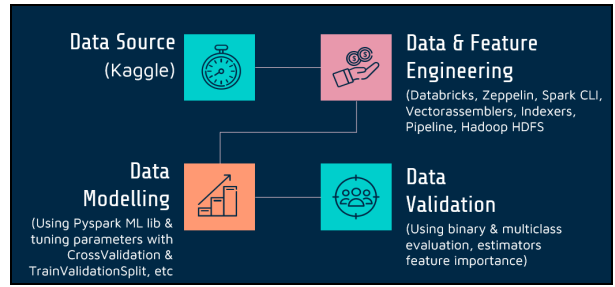

*Figure 1 – Architecture Diagram*

### 5. Implementation Flowchart

To begin with, the raw dataset, which has both files Books_rating.csv & books_data.csv, was downloaded from a trusted source (Kaggle). To perform prediction and build algorithms we used Machine Learning. The Machine Learning workflow has five steps as depicted in Figure 2.

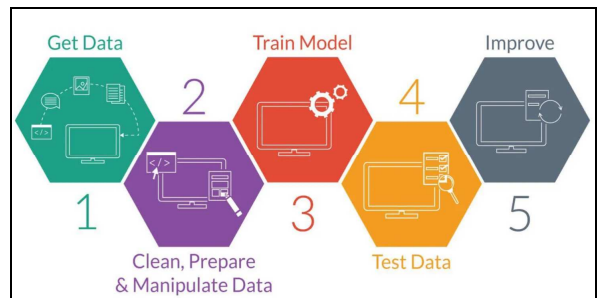

*Figure 2 – Machine Learning Workflow*

The first step is to Get Data the data, we downloaded the dataset from Kaggle. After that, we cleaned, prepared and manipulated data. Then we trained the model after which we tested the data and then we had multiple iterations to improve our model.

The implementation flow chart is shown below in Figure 3. The first and foremost step was to gather data in which we found out our project's objective and looked for data sources. The next stage was Data & Feature Engineering, here we prepared the data so as to use for modeling. Then we did the Data Split, wherein we had split the data and prepared it to train the model. In the next stage, which is Train, Test and Validate, we performed training, testing, and validating our model. Finally, in the Evaluation stage we evaluated the models that we built by using the measures for accuracy.

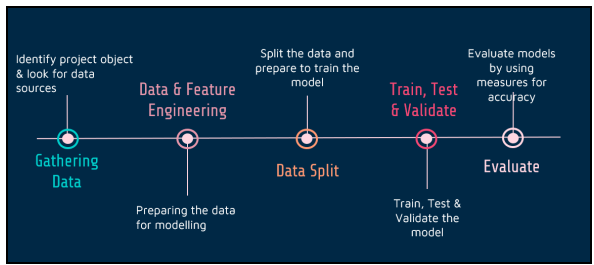

*Figure 3 - Implementation Flowchart*

### 6. Data Engineering

For our data set, we had two tables are as follows: Books_data (Table 3) and Books_Rating(Table 4)

*Table 3 – Books_data Table Description*

| Column Name | Data Type |
|---|---|
| title | string |
| description | string |
| authors | string |
| image | string |
| preview | string |
| publisher | string |
| publish_date | bigint |
| info_link | string |
| categories | string |
| ratings_count | int |

*Table 4 – Books_ratings Table Description*

| Column Name | Data Type |
|---|---|
| id | bigint |
| title | string |
| price | string |
| user_id | string |
| profile_name | string |
| r_helpfulness | string |
| r_score | int |
| r_time | bigint |
| r_summary | string |
| r_review | string |

### 7. Machine Learning

Our goal is to predict the rating score of Amazon books using various features such as price, review time, review summary, and review text. Initially, multiclass classification algorithms like Logistic Regression, Random Forest, and Decision Tree were employed, but the accuracy was found to be unsatisfactory. Consequently, the rating scores were transformed into a binary format, with scores 1-3 categorized as 0 and scores 4-5 as 1. Binary classification algorithms including GBT Classifier, Linear SVC, and Logistic Regression were then applied, along with techniques like TrainValidationSplit and CrossValidation for model building.

The below Figure 4 shows the various Machine Algorithms and the Recommendation model that were used in our project.

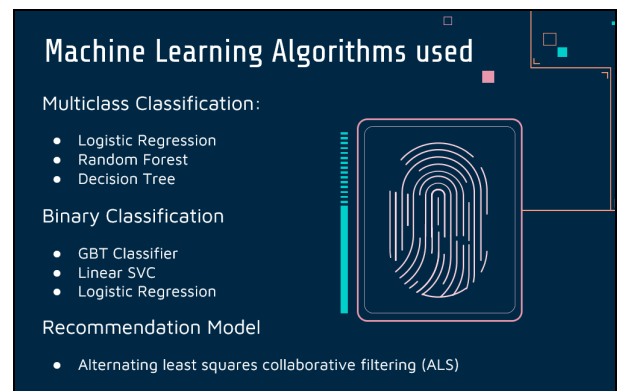

*Figure 4 – Machine Learning Algorithms and Recommendation Model details*

For the Feature Importance refer the table 5.

*Table 5 – Feature Importance*

| Feature | Importance |
|---|---|
| normFeatures2 | 0.208117 |
| numFeatures1 | 0.155013 |
| features2 | 0.083177 |
| featuresl | 0.026313 |

To handle the different data types in the features, the numeric features ("Price" and "review/time") were processed using MinMaxScaler, while the text feature ("review/summary") underwent preprocessing steps such as CountVectorizer, Tokenizer, StopWordsRemover, and IDF. The results indicated that transforming the scores to binary format led to higher prediction accuracy.

Comparing the multiclass classification algorithms, Logistic Regression outperformed the others in terms of both accuracy and F1 score, while requiring less time for model building. In the binary classification task, GBT Classifier had a lower accuracy and F1 score compared to Linear SVC and Logistic Regression, and it also took significantly more time to train. Linear SVC and Logistic Regression exhibited similar performance in terms of accuracy and F1 score, but Linear SVC had a slightly higher accuracy.

*Table 6 – Comparison of Models for Multiclass Classification*

| Model Name | Accuracy | Precision | Recall | F1 | Time |
|---|---|---|---|---|---|
| Decision Tree (tvs) | .63 | .57 | .63 | .53 | 4.7h |
| Random Forest (tvs) | .61 | .37 | .61 | .46 | 1.6h |
| Logistic Regression (tvs) | .64 | .58 | .64 | .59 | 44m |
| Decision Tree (cv) | .58 | .50 | .58 | .48 | 6.9h |
| Random Forest (cv) | .58 | .55 | .58 | .43 | 2.2h |
| Logistic Regression (cv) | .58 | .52 | .58 | .53 | 1.1h |

In addition to the classification models, recommendation models were also built using ALS and ALS Implicit techniques. ALS achieved an RMSE of 1.1 and an R2 of -1.2, while ALS Implicit resulted in an RMSE of 2.2 and an R2 of -10.6. These metrics provide insights into the performance of the recommendation models, with lower RMSE and higher R2 indicating better accuracy and fit to the data, respectively.

*Table 7 – Comparison of Models for Binary Classification*

| Model Name | Accuracy | Precision | Recall | F1 | Time |
|---|---|---|---|---|---|
| GBT Classifier (tsv) | .82 | .83 | .82 | .75 | 26h |
| Linear SVC (tsv) | .88 | .85 | .88 | .86 | 50m |
| Logistic Regression (tsv) | .85 | .84 | .85 | .83 | 48m |
| GBT Classifier (cv) | .79 | .80 | .79 | .73 | 30h |
| Linear SVC (cv) | .80 | .77 | .80 | .75 | 1.1h |
| Logistic Regression (cv) | .79 | .76 | .79 | .76 | 1h |

*Table 8 – Recommendation Models*

| Model | RMSE | R2 |
|---|---|---|
| ALS | 1.1 | -1.2 |
| ALS Implicit | 2.2 | -10.6 |

## 8. Conclusion

We aimed to predict Amazon book ratings by considering features like price, review time, review summary, and review text. Initially, we used multiclass classification algorithms (Logistic Regression, Random Forest, Decision Tree). We then transformed the ratings into binary format: 1-3 as 0 and 4-5 as 1. Binary classification algorithms (GBT Classifier, Linear SVC, Logistic Regression) were employed. We generalized models with TrainValidationSplit and CrossValidation. Numeric features were processed to improve the accuracy with MinMaxScaler, while the text feature underwent preprocessing (CountVectorizer, Tokenizer, StopWordsRemover, IDF). Binary format improved prediction accuracy. Logistic Regression performed best among multiclass classifiers. In binary classification, Linear SVC and Logistic Regression had similar accuracy and F1 score, while Linear SVC has slightly higher accuracy. Recommendation models (ALS, ALS Implicit) were also built. ALS had an RMSE of 1.1 and R2 of -1.2, while ALS Implicit had an RMSE of 2.2 and R2 of -10.6. Lower RMSE and higher R2 indicate better accuracy and fit for recommendation models, respectively.